\documentclass[twocolumn,showpacs,preprintnumbers,amsmath,amssymb,pra]{revtex4}

\usepackage{latexsym,epsfig,amssymb,amsfonts,amsmath,graphicx}

\newcommand{\ket}[1]{\left | \, #1 \right \rangle}

\newcommand{\ii}{\mathrm{i}}
\newcommand{\eh}{\mathrm{e}}
\newcommand{\diff}{\mathrm{d}}

\newcommand{\av}[1]{\langle #1\rangle}

\newcommand{\figref}[1]{Fig.~\ref{#1}}

\begin{document}

\title{Optical lattice quantum Hall effect}

\author{Rebecca N. Palmer$^{1,2}$, Alexander Klein$^{1,3}$, and Dieter
Jaksch$^{1,3}$}

\affiliation{$^1$ Clarendon Laboratory, University of Oxford, Parks
Road, Oxford OX1 3PU, United Kingdom \\
$^2$ School of Physics and Astronomy, University of Leeds, Leeds LS2
9JT, United Kingdom \\
 $^3$ Keble College, Parks Road, Oxford OX1 3PG, United Kingdom}

\begin{abstract}
  We explore the behavior of interacting bosonic atoms in an
  optical lattice subject to a large artificial magnetic field. We
  extend earlier investigations of this system where
  the number of magnetic flux quanta per unit cell $\alpha$ is close
  to a simple rational number [Phys.~Rev.~Lett.~{\bf 96}, 180407 (2006)].
  Interesting topological
  states such as the Laughlin and Read-Rezayi states can occur even
  if the atoms experience a weak trapping potential in one
  direction. An explicit numerical calculation near $\alpha =
  1/2$ shows that the system exhibits a striped vortex lattice
  phase of \emph{one} species, which is analogous to the behavior
  of a \emph{two}-species system for small $\alpha$.
  We also investigate methods
  to probe the encountered states. These include spatial correlation
  functions and the measurement of noise correlations in time-of-flight
  expanded atomic clouds. Characteristic differences arise which
  allow for an identification of the respective quantum Hall states.
  We furthermore discuss that a
  counterintuitive flow of the Hall current occurs for certain
  values of $\alpha$.
\end{abstract}
\date{\today}
\pacs{37.10.Jk,03.75.Lm,73.43.-f}


\maketitle


\section{Introduction}

Since their discovery, the integer \cite{vonKlitzing-PRL-1980} and
fractional quantum Hall effects \cite{Tsui-PRL-1982,Tsui-PRB-1982}
have attracted much attention on the theoretical as well as on the
experimental side. They were first observed in semi-conductor
samples which confine the electrons in a quasi two-dimensional layer
and are characterized by the ratio $\nu$ of particles to magnetic
flux quanta penetrating the layer. For the integer quantum Hall
effect, this ratio is an integer number, whereas the fractional
quantum Hall effect exhibits a simple rational (non-integer) $\nu$
and has interparticle interactions as an essential component. Some
of the excitations of the fractional quantum Hall effect are
promising candidates for topological quantum computing
\cite{Kitaev-AP-2003}, since they are separated from the ground
state by an energy gap and they might exhibit fractional, anyonic
statistics \cite{Moore-NPB-1991,Bravyi-PRA-2006,DasSarma-RMP-2008}.

Despite the work which has been put into this field, the mechanisms
which lead to the Hall effects are not completely understood
\cite{Murthy-RMP-2003,DasSarma-RMP-2008}. Especially theoretical
investigations are hampered by the complexity of the system: A
many-body quantum description is necessary to capture the relevant
effects such as the strong correlation of the electrons, which means
that the calculational effort scales exponentially with the number
of particles to be described. To gain insight into the behavior of
the quantum Hall effects it is therefore worthwhile to investigate
whether these phenomena occur in alternative systems as well.

Ultracold atoms in optical lattices are a system where the quantum
Hall effect can indeed be observed. As some of the present authors
have shown in an earlier publication \cite{Palmer-PRL-2006}, for a
suitable choice of the external parameters various topological
states occur in these systems such as the Laughlin state
\cite{Laughlin-PRL-1983} or the Read Rezayi states
\cite{Read-PRB-1999}. These states have been derived by developing a
model that is valid near simple rational numbers of magnetic flux
quanta per lattice cell, $\alpha$. We showed that the occurring
states can be distinguished by shot noise and Hall current
measurements, where in the latter case unexpected sign changes can
occur.

In the present paper we extend these investigations. We will give
details of how to expand the Hamiltonian describing cold bosonic
atoms in an optical lattice subject to an artificial magnetic field
for the case that $\alpha$ is close to a simple rational $\alpha_c =
l/n$ with $l$ and $n$ small integers. We will also give more insight
into how to distinguish the occurring states by noise correlation or
Hall current measurements. In addition to these extensions, we focus
on other phenomena occurring in our setup. By using a Gutzwiller
ansatz we demonstrate that for $\alpha$ close to $\alpha_c = 1/2$
and large filling fractions a two-component-like striped vortex
state occurs which shows that the analytical description derived in
\cite{Palmer-PRL-2006} is meaningful. The presence of the magnetic
field leads to an additional small scale structure for values of
$\alpha$ close to $\alpha_c$, which can be made visible in
time-of-flight expansions as we will illustrate for $\alpha_c =
1/2$.

The applicability of our scheme is ensured by the experimental and
theoretical successes in the field of optical lattices during the
last couple of years. Theoretical investigations have shown that a
whole wealth of models can be simulated \cite{Lewenstein-AiP-2006},
such as the Hubbard Hamiltonian for bosons, fermions, or mixtures
\cite{Jaksch-PRL-1998,Jaksch-Ann-2005}, spin-spin interactions
\cite{Sorensen-PRL-1999,Duan-PRL-2003}, high-temperature
superconductivity \cite{Hofstetter-PRL-2002,Klein-PRA-2006b}, or the
formation of polarons
\cite{Bruderer-PRA-2007,Klein-NJP-2007,Bruderer-NJP-2008}.
Theoretical investigations show that these models exhibit rich phase
diagrams which should be accessible with current experimental
techniques \cite{Mathey-PRL-2004, Lewenstein-PRL-2004,
Mehring-PRA-2008, Duan-PRL-2003, Buchler-PRL-2003}. The necessary
low temperatures can be achieved by appropriate cooling methods
\cite{Popp-PRA-2006,Popp-NJP-2006,Griessner-PRL-2006,Griessner-NJP-2007}.
It is also possible to implement artificial magnetic fields in this
setup, for example by rotating the lattice
\cite{Cooper-PRL-2001,Paredes-PRL-2001,Polini-PRL-2005,Polini-LP-2004}.
This comes, however, with the disadvantage that a centrifugal term
occurs, which has to be balanced by an external trapping potential.
To circumvent this problem alternative schemes have been developed,
exploiting Raman assisted hopping
\cite{Jaksch-NJP-2003,Mueller-PRA-2004}, oscillating quadrupole
fields \cite{Sorensen-PRL-2005}, or laser beams with orbital angular
momentum \cite{Juzeliunas-PRL-2004,Juzeliunas-PRA-2005,
Juzeliunas-PRA-2006}. Some of these schemes even exhibit non-Abelian
gauge potentials \cite{Osterloh-PRL-2005,Ruseckas-PRL-2005},
allowing for the investigation of new phenomena in the integer
quantum Hall effect \cite{Goldman-arXiv-2007} or the observation of
the non-Abelian Aharanov-Bohm effect \cite{Jacob-APB-2007}.

In the presence of a lattice, the strength of the artificial
magnetic field is suitably characterized by the number of effective
magnetic flux quanta per unit cell, $\alpha$, which for a
conventional semiconductor setup is typically very small, $\alpha
\ll 1$. In the optical lattice setup, however, it is possible to
achieve values close to $\alpha = 1/2$, a regime which is almost
impossible to achieve in the condensed matter case due to the
required huge magnetic fields. This allows for the experimental
investigation of so far unobserved phenomena such as the Hofstadter
butterfly
\cite{Hofstadter-PRB-1976,Jaksch-NJP-2003,Geisler-PRL-2004,Umucalilar-PRA-2007},
a fractal energy spectrum originally predicted for electrons subject
to a large magnetic field. It is also worthwhile to investigate the
energy gap of anyonic excitations near $\alpha = 1/2$. If it is
sufficiently large, it would make topological quantum computing more
feasible since thermally excited anyons are the major source of
errors in these schemes \cite{Kitaev-AP-2003}.

There have been previous investigations of ultracold atoms in
optical lattices subject to large artificial magnetic fields or
rotation. For instance, in \cite{Ho-PRL-2002} a rotation scheme of
an optical lattice was proposed which leads to quantum Hall and spin
liquid properties of a spin-1 boson cluster. The dependence of the
state on the rotation frequency has been investigated in
Ref.~\cite{Popp-PRA-2004}, where it was shown that one can change
the state adiabatically from a Gaussian to a Laughlin state.
Transport properties of one and two particles in a rotating lattice
have been investigated in Ref.~\cite{Bhat-PRA-2007} using a Kubo
formalism. Recently, the influence of boundary conditions on the
Hall states was investigated and Chern numbers were employed to
characterize the signature of the topological states
\cite{Hafezi-PRA-2007,Hafezi-EPL-2008}. These authors also discussed
that the quantum Hall states can be made more stable and the energy
gap to excitations can be increased by using atoms with a strong
dipole-dipole interaction such as chromium \cite{Hafezi-PRA-2007}.

An experimental implementation of the proposed methods and the
observation of the predicted effects should be possible in the near
future \cite{Jaksch-Ann-2005}. The creation of fast rotating traps
has already been demonstrated in
Refs.~\cite{Bretin-PRL-2004,Schweikhard-PRL-2004}, which have
allowed one to investigate effects such as vortex pinning
\cite{Tung-PRL-2006}. Loading an optical lattice with a well-defined
number of atoms, thereby allowing one to tune $\nu$ independently
from the strength of the artificial magnetic field, has been
achieved by exploiting spin oscillations
\cite{Thalhammer-PRL-2006,Widera-PRA-2005,Gerbier-PRA-2006}, and the
number statistics of ultracold atom systems has been examined in
\cite{Oettl-PRL-2005,Gerbier-PRL-2006}. Also, measurement schemes
have been implemented which allow for an investigation of second
order correlation functions such as noise correlations
\cite{Foelling-Nature-2005}, thereby revealing additional quantum
properties of the states.

Our paper is organized as follows. In Sec.~\ref{Sec:Hamiltonian}, we
introduce the Hamiltonian which describes the atoms in the optical
lattice subject to an artificial magnetic field in Landau gauge. For
weak magnetic fields, a continuum approximation is applied as shown
in Sec.~\ref{Sec:Small_alph}. We extend our investigations in
Sec.~\ref{sec:high-field} to higher artificial magnetic fields, such
that $\alpha$ is close to simple rational values where the system is
well-described by multi-component continuous wave functions. In
Sec.~\ref{Sec:Probing} we present how the different quantum Hall
states can be distinguished from each other. The used methods
include time-of-flight expansions, measuring noise correlations in
the time-of-flight expanded clouds, and observing the occurring Hall
mass current. We conclude in Sec.~\ref{sec:fqh-conclusions}.


\section{Hamiltonian\label{Sec:Hamiltonian}}

Throughout this work we assume a two-dimensional optical lattice
\cite{Jaksch-PRL-1998,Jaksch-Ann-2005}, which can be experimentally
achieved for example by using a three-dimensional lattice with the
hopping into one direction strongly suppressed
\cite{Koehl-JLTP-2005}. The atoms are then confined in
two-dimensional planes, of which we consider only one in the
following. The artificial magnetic field term is created using one
of the several methods discussed in the Introduction
\cite{Jaksch-NJP-2003,Mueller-PRA-2004,Sorensen-PRL-2005,
Juzeliunas-PRL-2004,Juzeliunas-PRA-2005,Juzeliunas-PRA-2006}. We
furthermore assume that the total potential energy experienced by
the atoms is given by $V(p,q)$, where $p,q$ label the lattice sites
in the $x$- and $y$-directions, respectively. In Landau gauge, which
is more convenient for our purposes, the Hamiltonian describing the
atoms is given by
\cite{Palmer-PRL-2006,Jaksch-NJP-2003,Sorensen-PRL-2005}
\begin{equation} \label{Eq:Hamiltonian}
\begin{split}
  \hat H =& - \sum_{p,q} \left( J_x \eh^{2\pi \ii \alpha q} \hat a^\dagger_{p,q}
  \hat a_{p-1,q}
       + J_y \hat a^\dagger_{p,q}\hat a_{p,q-1} + \mathrm{H.c.} \right)\\
     &+V(p,q)\hat a^\dagger_{p,q}\hat a_{p,q}
       +\frac{U}{2}\hat a^\dagger_{p,q}\hat a^\dagger_{p,q}\hat a_{p,q}\hat a_{p,q} \,,
\end{split}
\end{equation}
where the magnetic field strength is parameterized by the number of
flux quanta per lattice cell $\alpha$. The operator $\hat
a^\dagger_{p,q}$ creates an atom in lattice site $(p,q)$, $J_x$ and
$J_y$ are the hopping constants in the $x$ and $y$ directions,
respectively, and $U$ describes the strength of the on-site
interaction between the atoms. Due to the artificial magnetic field
terms, the atoms acquire a phase $2\pi\alpha$ when they hop around a
lattice cell once. Since the value of $\alpha$ is only defined
modulus one, we may restrict it to $0\leq\alpha <1$.

It is instructive to compare our setup to other systems in which the
quantum Hall effect can be observed, such as a semiconductor
structure or a rotating ultracold atomic gas without an optical
lattice. The main characteristics and typical values are shown in
Table \ref{tab:comparison}. Due to the small lattice spacing $d \sim
10^{-10}\mathrm{m}$ in the original solid state semiconductor
systems, for instance, Si or GaAs, typically only small values
$\alpha = e B d^2/ 2 \hbar \pi \sim 10^{-4}$ can be achieved with
available magnetic fields $B$. Here, $e$ is the charge of the
electron. Although this can be overcome by using superlattice
structures \cite{Albrecht-PRL-2001,Geisler-PRL-2004}, defects or
impurities are inevitable in a real crystal. In contrast, the
optical lattice setup allows for a virtually defect-free
implementation of the quantum Hall Hamiltonian, and values up to
$\alpha \sim 1$ can be achieved
\cite{Jaksch-NJP-2003,Sorensen-PRL-2005,Mueller-PRA-2004,Hafezi-PRA-2007}.
Comparing the cyclotron frequency $\Omega = \pi \alpha \hbar / \bar
m d^2$ (for a definition of the effective mass $\bar m$ in the
lattice setup see the next section) to the temperatures which can be
achieved in the respective realizations shows that similar regimes
can be realized with all three methods when appropriate cooling
techniques are exploited in the atom setup
\cite{Popp-PRA-2006,Popp-NJP-2006,Griessner-PRL-2006,Griessner-NJP-2007}.
Low ratios of temperature over magnetic field are important to avoid
thermal excitations which would spoil the applicability of the
system for topological quantum computation \cite{DasSarma-RMP-2008}.

In contrast to the semiconductor scheme, the two atomic realizations
provide more opportunities for tunability and probing the states.
Both atom setups allow for additional measurements to be carried
out, such as time-of-flight expansions and measuring second order
correlation functions \cite{Altman-PRA-2004,Foelling-Nature-2005}.
The lattice setup furthermore enables an easy tuning of the system
parameters such as the hopping or the interaction between the atoms
\cite{Jaksch-Ann-2005}. In a standard optical lattice setup, this
interaction is given by a contact term $\delta(r)$. By using certain
atomic species such as chromium one can also implement dipole-dipole
interactions which might stabilize the quantum Hall states
\cite{Hafezi-PRA-2007}. It is also possible to extend the contact
interaction by immersing the optical lattice into a Bose-Einstein
condensate, where an additional off-site interaction occurs
\cite{Bruderer-PRA-2007,Klein-NJP-2007,Klein-PRA-2005}.

Compared to the rotating setup, in a lattice there exist methods
which do not involve a centrifugal term that has to be balanced by
an additional external potential. Although experiments with rotating
traps already exist
\cite{Bretin-PRL-2004,Schweikhard-PRL-2004,Tung-PRL-2006}, reaching
the quantum Hall regime is challenging due to the need of balancing
the centrifugal force.

\begin{table*}
\begin{tabular}{|l|c|c|c|}
\hline
System&Electrons (GaAs)&Rotating gas (Rb)&Lattice (Rb)\\
\hline
Particles&Electrons&Atoms&Atoms\\
Statistics&Fermi&Bose&Bose\\
Interaction potential&$1/r$&$\delta(r)$&$\delta(r)$\\
Confinement&Sharp edges&Smooth trap&Smooth trap\\
\hline
Physical parameters&&&\\
\hline
System diameter&$\sim 10^{-2}\rm m$&$10^{-5}-10^{-4}{\rm m}$&$10^{-5}-10^{-4}{\rm m}$\\
Number of particles $N$&$\sim 10^{11}$&$10^5-10^6$&$\sim 10^4$\\
2D number density $\varrho$&$1-2\times 10^{15}{\rm m}^{-2}$&$\sim 10^{12}{\rm m}^{-2}$&$10^{10}-10^{11}{\rm m}^{-2}$\\
Cyclotron frequency (CF) $\Omega$&$\sim 10^{14}{\rm Hz}$&$10^2-10^3{\rm Hz}$&$\sim 10^3{\rm Hz}$\\
Free mass $m_0$&$9\times 10^{-31}{\rm kg}$&$1.5\times 10^{-25}{\rm kg}$&$1.5\times 10^{-25}{\rm kg}$\\
Effective mass $m$&$6\times 10^{-32}{\rm kg}$&$1.5\times 10^{-25}{\rm kg}$&$10^{-24}-10^{-23}{\rm kg}$\\
Lattice spacing $d$&$2\times 10^{-10}{\rm m}$&n/a&$\sim 10^{-6}{\rm m}$\\
Temperature $T$&$0.1-0.3{\rm K}$&$\sim 10^{-9}{\rm K}$&$\sim 10^{-11}{\rm K}$\\
\hline
Dimensionless parameters& & & \\
\hline
$\alpha$ (flux quanta/unit cell)&$10^{-4}$&n/a&$0.1-0.5$\\
$\nu$ (particles/flux quanta)&$0.2-10$&$10^3-10^4$&$1-10$\\
$\varrho u/\hbar\Omega$ (interaction/CF)& n/a & $0.1-1$&$\sim 0.1$\\
$k_BT/\hbar\Omega$ (temperature/CF)&$10^{-3}$&$10^{-1} - 1$&$\sim 10^{-3}$\\
\hline
Internal states&spin or bilayer&hyperfine&hyperfine\\
\hline
Tunability&&&\\
\hline
$\Omega,\varrho,T$&yes&yes&yes\\
$um$&no&yes&yes\\
$m_x/m_y$&no&no&yes\\
\hline
\end{tabular}
\caption{\label{tab:comparison}Comparison of quantum Hall systems.
The regime of interest is $\nu\sim 1$, and $\hbar\Omega\gg\varrho
u\gg k_\mathrm{B} T$. The values for the BEC are from current
experiments \cite{Bretin-PRL-2004,Schweikhard-PRL-2004}, which are
still slightly outside this regime. The low temperatures for the
lattice system can be achieved by using, for instance, the methods
in
Refs.~\cite{Popp-PRA-2006,Popp-NJP-2006,Griessner-PRL-2006,Griessner-NJP-2007}.
For the lattice, the data shown is for a single plane only. If the
two-dimensional plane is realized using a three-dimensional lattice
with the hopping in one direction strongly suppressed, the quantum
Hall effect occurs independently in each plane and multiple planes
can be used to increase the signal strength without increasing
$\nu$.}
\end{table*}


\section{Small $\alpha$ limit: continuum approximation\label{Sec:Small_alph}}

In earlier publications \cite{Sorensen-PRL-2005,Palmer-PRL-2006}, it
was shown that for small values of $\alpha$ the influence of the
optical lattice is negligible and we can approximate the lattice gas
by a continuous wave function. In the present section, we give a
more detailed account of this analysis. We first consider a single
particle in the optical lattice described by the discrete wave
function $\ket{\psi}=\sum_{p,q} \psi(p,q) \hat
a^\dagger_{p,q}\ket{0}$, which obeys the normalization condition
$\sum_{p,q} |\psi(p,q)|^2 = 1$. The Hamiltonian
Eq.~(\ref{Eq:Hamiltonian}) acts on this wave function as
\begin{equation}
\begin{split}
  \hat H \ket{\psi} =& \sum_{p,q} \{-J_x[\eh^{2\pi \ii \alpha q}
  \psi(p-1,q)+\eh^{-2\pi \ii \alpha q} \psi(p+1,q)]  \\
  &\quad -J_y[\psi(p,q-1)+\psi(p,q+1)]          \\
  &\quad +V(p,q)\psi(p,q)\}a^\dagger_{p,q}|0\rangle \,.
\end{split}
\end{equation}
For $\alpha \ll 1$ and a weak trapping potential $V(p,q)$, the wave
function varies only slowly from one lattice site to the neighboring
ones, and we can approximate the state by a continuous wave function
$\phi(x,y)$, where $\psi(p,q)=d \phi(pd,qd)$ with $d$ the lattice
spacing and $\int \! |\phi(x,y)|^2 \,\diff x \diff y = 1$. The
dynamics of this wave function is governed by the Hamiltonian
\begin{equation}
\begin{split}
  H_0 =& - J_x \left[2 - \frac{d^2}{\hbar^2} \left(\ii \hbar \frac{\partial}{\partial x} -
  \frac{2 \pi \alpha \hbar y}{d^2} \right)^2 \right]  \\
  &- J_y
  \left[ 2 -\frac{d^2}{\hbar^2} \left( \ii \hbar \frac{\partial}{\partial y}\right)^2
   \right] + V(x,y)
  \,.
\end{split}
\end{equation}
This Hamiltonian can be transformed into a more familiar form by
defining the effective masses $m_{x,y} = \hbar^2/2 J_{x,y} d^2$,
$\bar m = \sqrt{m_x m_y}$, and the cyclotron frequency $\Omega = \pi
\alpha \hbar / \bar m d^2$. After discarding a constant energy term
we get
\begin{equation}\label{eq:cont-1par}
  H_0 =  \frac{1}{2 m_x} \left(  \ii \hbar
  \frac{\partial}{\partial x} - 2 \bar m \Omega y \right)^2
  - \frac{\hbar^2}{2 m_y} \frac{\partial^2}{\partial y^2} + V(x,y)
  \,.
\end{equation}
This is the familiar single particle quantum Hall Hamiltonian with
an artificial ``electric'' field potential $V(x,y)$ and anisotropic
mass, which can be redefined into an anisotropic magnetic length.
For deriving the energy levels we assume that the potential $V(x,y)$
is constant in the $x$ direction and forms a harmonic potential in
the $y$ direction, $V(x,y) = m_y \omega^2 y^2/2$. The Hamiltonian is
then translational invariant in the $x$ direction, which justifies
the ansatz $\phi(x,y) = \exp(-\ii K x) F(y)$. The Hamiltonian $H_0$
acts on this wave function as
\begin{equation}
  H_0 \phi = \left[- \frac{\hbar^2}{2m_y} \frac{\partial^2}{\partial y^2} +
  \frac{1}{2} m_y \omega_\mathrm{eff}^2 (y - y_c)^2
  +
  \frac{\omega^2 \hbar^2 K^2}{2 m_x \omega_\mathrm{eff}^2}\right] \phi \,,
\end{equation}
where $\omega_\mathrm{eff} = \sqrt{4 \Omega^2 + \omega^2}$, and $y_c
= 2 \Omega \hbar K / \omega_\mathrm{eff}^2 \bar m$. This expression
describes a displaced harmonic oscillator with an energy offset
depending on $K$. For $\omega = 0$ we retain the usual Landau
levels, whereas for $\omega \neq 0$ the degeneracy within one Landau
level is lifted and the energies are given by
\begin{equation}
  E = \left(n_{\mathrm{LL}} + \frac{1}{2} \right) \hbar\omega_\mathrm{eff} +
  \frac{\omega^2 \hbar^2 K^2}{2 m_x \omega_\mathrm{eff}^2} \,,
\end{equation}
where $n_{\mathrm{LL}} \geq 0$ is an integer.

The continuum approximation can be extended to the case of more than
one particle in the lattice, provided that the wave function still
varies slowly over the distance $d$ between two lattice sites. To
achieve this, the average distance between the particles needs to be
large compared to $d$, which restricts the particle density to
$\varrho \ll 1/d^2$. In this case, the continuum Hamiltonian is
given by
\begin{equation} \label{Eq:ContAppIA}
  H = \sum_j H_0(x_j,y_j) + \frac{u}{2} \sum_{i,j} \delta(x_i - x_j)
  \delta(y_i - y_j) \,,
\end{equation}
where $u = U d^2$. This Hamiltonian describes the quantum Hall
effect with a contact interaction term, in contrast to the typically
screened $1/r$ interaction of the Coulomb potential for electrons in
solids. In analogy to the solid state fractional quantum Hall effect
we define the filling factor $\nu=\hbar\varrho\pi/\bar
m\Omega=\varrho d^2/\alpha$. Since, in this section, we are
interested in the limits $\alpha \ll 1$ and $\varrho \ll 1/d^2$ we
will hence find the same states as in a continuum bosonic quantum
Hall system, for example, the $\nu=1/2$ Laughlin state, the
$\nu=1,3/2,\ldots$ Read-Rezayi states, or a vortex lattice
\cite{Palmer-PRL-2006,Read-PRB-1999,Cooper-PRA-2005}.

The strength of the interparticle interaction $u$ plays a crucial
role if the setup is to be used for topological quantum computing.
For this, anyonic quasiholes are created and they are moved around
each other using focussed lasers, thereby inducing qubit operations
\cite{Paredes-PRL-2001}. However, for the Read-Rezayi (non-Abelian
anyon) states to appear the interaction must be weak compared to the
spacing between the Landau levels given by $2 \hbar \Omega$, to
avoid excitations to higher levels. This reduces the
quasiparticle-quasihole pair creation gap $\Delta_g \sim um \Omega/2
\pi \hbar$, where we assume $m \equiv \bar m = m_x = m_y$. As
thermally created anyons moving around the computational anyons are
a source of error, such computation would require low temperatures
$T \ll u m \Omega /2 \pi \hbar k_\mathrm{B} \ll 2 \hbar \Omega /
k_\mathrm{B}$, which for typical experimental setups are on the
order of a few 10\,nK. These temperatures might be reached using the
methods of
\cite{Popp-PRA-2006,Popp-NJP-2006,Griessner-PRL-2006,Griessner-NJP-2007}.
Alternatively, it has been suggested that non-Abelian field quantum
Hall states \cite{Lewenstein-AiP-2006,Osterloh-PRL-2005} may offer
non-Abelian excitations in the lowest density state, allowing a
strong interaction to be used to increase the gap to $\sim \hbar
\Omega$, but this has yet to be confirmed.


\section{\label{sec:high-field}Near simple rational $\alpha$:
  multi-component wavefunctions}

We will now relax the condition of small $\alpha$ and investigate
the properties of the quantum gas for an $\alpha$ close to simple
rational values $\alpha \approx l/n$, where $l$ and $n$ are small
integers. The wave function can then be approximated by a set of $n$
smooth, slowly varying functions which correspond to $n$ different
components of the gas. These components should not be confused with,
for instance, different internal states of the atoms. They rather
correspond to different small scale structures of the gas. Based on
previous work \cite{Palmer-PRL-2006}, we will give details on how to
derive these expressions and consider some special cases. We
especially calculate the ground state wave function for a set of
representative parameters numerically, which confirms that for
$\alpha$ close to 1/2 the two-component description is meaningful.
For simplicity, in the following we will consider the case of
isotropic hopping, that is, $J_x = J_y \equiv J$, leading to
effective masses $m_x = m_y = \bar m \equiv m$.

\subsection{\label{sec:high-field-1}Single particle states}

As discussed in \cite{Palmer-PRL-2006}, numerical calculations for
simple $\alpha = l/n$ and weak external potentials $V$ suggest that
the single particle ground state functions exhibit an $n$-site
periodic pattern superimposed on a smooth large-scale variation.
This motivates the representation $\psi(np+i,nq+j)=d \sum_k \chi_k(d
(n p +i) ,d (n q +j)){\mathbf v}^{(k)}_{ij}$, where $\chi_k$ is a
continuous, slowly varying function and ${\mathbf v}^{(k)}$ an
$n\times n$ matrix describing the small scale structure of the
atomic gas. We find by expansion about $\alpha_c\equiv l/n$ (see
Appendix \ref{app:high-field} for details) that there are $n$
degenerate matrices ${\mathbf v}^{(k)}$ of the form ${\mathbf
v}^{(k)}_{pq}=e^{2\pi ipk/n}{\mathbf v}_{q+k}$, where ${\mathbf v}$
is a fixed $n$-component vector for each $l,n$ and the subscript
$q+k$ wraps around mod $n$. Furthermore, the $\chi_k$ obey the
condition
\begin{align}\label{eq:chi-1par}
  &-\frac{C}{2m}\frac{\hbar^2\partial^2\chi_k}{\partial y^2}   \nonumber
  +\frac{C}{2m}\left(2m\tilde\Omega y
  -i\hbar \frac{\partial}{\partial x}\right)^2\chi_k
  +V(x,y)\chi_k  \\
  &\qquad
  =\left(E-\frac{E_0}{md^2}\right)\chi_k \,,
\end{align}
where $\tilde\Omega \equiv \hbar (\alpha-\alpha_c)\pi/(md^2)$ and
$C,E_0$ depend only on $l,n$.  This formula reduces to
Eq.~(\ref{eq:cont-1par}) for $\alpha_c=0/1$, and agrees well with
numerical calculations near other simple $\alpha_c$, especially
$\alpha_c = 1/2$ \cite{Palmer-PRL-2006}.

For this procedure to be consistent, the length scale $l_\chi$ over
which $\chi_k$ varies, which for a harmonic trap
$V(x,y)=\frac{1}{2}m\omega^2y^2$ is given by $l_\chi = (\hbar^2
C/(4C m^2\tilde\Omega^2 + m^2\omega^2))^{1/4}$, must be large
compared to the ``small scale'' periodicity $nd$, but small enough
that $2m\tilde\Omega l_\chi d/\hbar\ll 1$ (cf.~Appendix
\ref{app:high-field}), that is
\begin{equation}
\frac{1}{n d} \gg \frac{1}{l_\chi} \gg \frac{2 m d
\tilde\Omega}{\hbar} \,,
\end{equation}
or equivalently
\begin{equation}
\frac{1}{n}\gg \left[ \frac{\pi^2}{C}(4C (\alpha-\alpha_c)^2 +
\beta^2)\right]^{\frac{1}{4}} \gg 2 \pi
(\alpha-\alpha_c)\label{eq:chi-validrange} \,.
\end{equation}
Here $\beta = md^2\omega / \hbar \pi$ is the dimensionless trap
strength. The first inequality shows that the range over which the
multi-component ansatz is valid gets narrower for larger
denominators $n$ of $\alpha_c$. We also see that the presence of the
trap is important to ensure the validity of the second inequality. A
stronger trap makes it easier to fulfill the second condition,
however, a too strong trap will ultimately lead to a contradiction
with the first condition. Indeed, in Ref.~\cite{Palmer-PRL-2006} it
was shown that for small $n$ the above wave functions compare very
well with numerical exact calculations and overlaps of more than
$99\%$ can be achieved for $\alpha \approx \alpha_c$ and
appropriately strong traps.

\subsection{Interacting particles}

Eq. (\ref{eq:chi-1par}) is extended to the many-particle case as
\begin{eqnarray}
  H&\approx&\int \! \diff x \diff y \, \sum_k  \chi_k^\dagger(x,y)
  \Bigg\{-\frac{C \hbar^2 }{2m}\frac{ \partial^2}{\partial y^2}\nonumber\\
  &&\quad
  +\frac{C}{2m}\left(2m\tilde\Omega y-
  i\hbar \frac{\partial}{\partial x}\right)^2+V(x,y)\Bigg\}
  \chi_k(x,y)\nonumber\\
  &&\quad + u \sum_{k_1,k_2,k_3,k_4}G_{k_1,k_2,k_3,k_4}
  \chi_{k_1}^\dagger(x,y)\chi_{k_2}^\dagger(x,y) \nonumber\\
  &&\qquad \times\chi_{k_3}(x,y)
  \chi_{k_4}(x,y)+\frac{E_0}{md^2} \,,\label{eq:chi-manypar}
\end{eqnarray}
where $G_{k_1,k_2,k_3,k_4}\equiv\sum_j {\mathbf v}_{j+k_1}{\mathbf
v}_{j+k_2}{\mathbf v}_{j+k_3}{\mathbf v}_{j+k_4}/n$ if
$k_1+k_2\equiv k_3+k_4$ mod $n$, and 0 otherwise. The conservation
mod $n$ is due to the $x$ quasimomentum $2\pi k/nd$ carried by
${\mathbf v}^{(k)}$. To give an example, for $\alpha_c=1/2$ we find
$G_{1111}=G_{2222}=3/2$ and
$G_{1212}=G_{2121}=G_{1122}=G_{2211}=1/2$. Particles with the same
$k$ interact more strongly because their ${\mathbf v}^{(k)}$ are
peaked on the same sites.

For $\alpha_c=1/2$, a change of basis from $\chi_{1,2}$ to
$\chi_\pm=\chi_1\pm \ii \chi_2$ makes this effective Hamiltonian
analogous to a bilayer fractional quantum Hall system
\cite{DasSarma-1997}, with $\chi_\pm$ being the two ``layers''.
However, the interaction ratio is 1:2 with the ``interlayer''
interaction being the stronger one, while it is equal or weaker in
most other realizations of multicomponent fractional quantum Hall
states.

\subsection{Some special states}

For the case that $\alpha_c = 1/2$ the system is equivalent to a
two-component gas as argued in the previous section. It is
well-established that for an interacting two-component gas the
lowest Landau level state with highest density and zero interaction
energy is the so-called 221 state \cite{Ezawa-2000} defined by
\begin{equation}
\begin{split}
  &\phi_{221}(z_1,\ldots,z_{N/2},w_1,\ldots,w_{N/2}) \\
  &=\left[\prod\limits_{i>j}(z_i-z_j)^2\right]\left[\prod\limits_{i>j}(w_i-w_j)^2\right]
  \left[\prod\limits_{i,j}(z_i-w_j)\right] \\
  &\quad\times\exp\left(-\sum\limits_i |z_i|^2/4+|w_i|^2/4\right)
  \,,
\end{split}
\end{equation}
where there are $N/2$ particles in one component with coordinates
$z=(x+\ii y)/r_0$ and $N/2$ in the other component with coordinates
$w=(x+\ii y)/r_0$, and $r_0=[\hbar/2m \tilde\Omega]^{1/2}$. For the
lattice setup at $\alpha = 1/2$, the two components are replaced by
the two ``layers'', and the effective filling factor of this state
is defined with respect to $\tilde \Omega$ and given by $\tilde\nu
\equiv \hbar \varrho\pi/m\tilde\Omega = 2/3$. The state can be
extended to a general $\alpha_c = l/n$ with $\tilde \nu = n/(n+1)$.

Since this state has exactly zero interaction energy, it would be
the lowest step of the density profile in a slowly varying external
potential. Adding more particles to the system leads to a trade-off
between an increasing potential energy when the particle is added to
the ``edge'' of the 221 state and an increased interaction energy if
the particle is added at the center of the trap where the potential
energy is lowest. This eventually leads to a stepped density
profile. So far it was not possible to determine with certainty what
the next higher step looks like, but it might be either a
$\tilde\nu=1$ Read-Rezayi \cite{Read-PRB-1999} state or a
$\tilde\nu=4/3$ non-Abelian spin singlet (NASS) state
\cite{Ardonne-PRL-1999}. Depending on the energies of the states,
which are not exactly known, also both steps can occur. Higher
states might be the Read-Rezayi states with $\tilde\nu=3/2$,
$\tilde\nu=2$, and so on, eventually leading to a vortex lattice
\cite{Sinova-PRL-2002}.

It is still unknown where exactly the transition to this vortex
lattice phase occurs. The Lindemann criterion
\cite{Sinova-PRL-2002,Rozhkov-PRB-1996} gives the order of magnitude
estimate $\nu \sim 10$ for both one- and two-component systems,
whereas numerical studies of the single component system
\cite{Cooper-PRA-2005} give $\nu \sim 2-6$. We hence conjecture that
our two-``layer'' system has a similar critical $\tilde \nu$. To
investigate the vortex lattice at a large filling factor and $\alpha
$ close to $1/2$, we have done numerical simulations using imaginary
time propagation of the Gutzwiller ansatz \cite{Krauth-PRB-1992}
\begin{equation}
  \ket{\psi} = \prod_{i,j} \left[ \sum_{n = 0}^{n_\mathrm{max}}
  c_{i,j,{n}}  \left( \hat a_{i,j}^\dagger \right)^{n}
  \ket{\mathrm{vac}}\right] \,,
\end{equation}
where the state is specified by the complex numbers $c_{i,j,{n}}$.
This ansatz can describe both superfluid vortex lattice and Mott
insulator states, but in general not off-site correlated states such
as the fractional quantum Hall states.

\begin{figure}
\centering
\includegraphics[width=8cm]{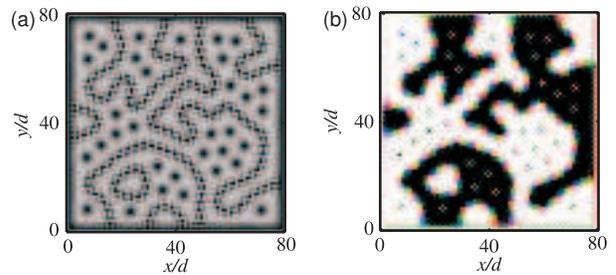}
\caption{\label{fig:vortices}(Color online) Numerical approximation
to the ground state of the $\alpha_c=1/2$ vortex lattice phase,
calculated using imaginary time propagation of a Gutzwiller ansatz.
(a) Density of the atom distribution, black stands for no density,
white means high superfluid density, red (hardly visible) means high
Mott insulator density. (b) Population of the two ``layers'', black
and white depict the two components, color encodes the phase of the
superposition of those two components. The parameters are
$\alpha=0.48$, $U=0.1J$, and $\tilde \nu = 21.2$.}
\end{figure}

A typical density distribution of our calculations using a random
initial state is shown in Fig.~\ref{fig:vortices}(a). It exhibits
alternating stripes as have been predicted for a continuum
two-component system \cite{Mueller-PRL-2002,Kasamatsu-PRL-2003}.
However, in our case the two components are not two different
species or states of one species, but represent the two ``layers''
corresponding to the two components of the wave function. For a
better distinction these two components are plotted with different
colors in Fig.~\ref{fig:vortices}(b). A direct measurement of these
two components analogous to the measurement of two different species
is not possible. Instead, to make the two components visible we
calculated the overlaps of the small scale solutions
$\mathbf{v}^{(k)}$ with groups of lattice sites of the Gutzwiller
solution. These overlaps correspond to the large-scale functions
$\chi_k$ at the respective lattice sites. Our calculations show that
the densities described by the two components indeed add up to the
total density of the system and confirm that at least for the chosen
parameters the two-component description is meaningful. We find that
the stripes straighten out only very slowly with increasing
imaginary time (or decreasing temperature), similar to the findings
in \cite{Kasamatsu-PRL-2003}, with a slight tendency to phase
separation. The perfectly straight stripes observed in
\cite{Mueller-PRL-2002} exist only in the zero temperature limit. At
exactly $\alpha=\alpha_c = l/n$, we have $\tilde\nu=\infty$ and one
has an $n$ component superfluid with no vortices. This may be the
possible superfluid phases found at $\alpha=1/3,1/2$ in the exact
diagonalizations of S{\o}rensen and coworkers
\cite{Sorensen-PRL-2005}.

\section{Probing the states\label{Sec:Probing}}

After having demonstrated that a very wide range of different states
can occur within our setup, the question arises how these states can
be distinguished from each other. Fractional quantum Hall states for
which Eq.~(\ref{eq:chi-manypar}) is valid are characterized by a
fixed $\tilde\nu$, and hence the density for these states will obey
the condition $\varrho\propto|\alpha-\alpha_c|$. For ``ordinary''
fractional quantum Hall states with $\alpha \ll 1$ the
characteristic parameter is $\nu$, which consequently leads to a
density that obeys $\varrho \propto \alpha$ as discussed in
Sec.~\ref{Sec:Small_alph}. In contrast, for Mott insulating states,
the density is fixed to integer filling and does neither depend on
$\alpha$ nor on $\alpha - \alpha_c$. An in-trap density profile
measurement could hence distinguish these states by comparing
measurements at different artificial magnetic fields (and thus
different $\alpha$), as well as measuring the ratio $\nu$ (or
$\tilde \nu$), at which these incompressible states occur.

However, a more clear-cut distinction between the occurring states
is possible by using methods such as time-of-flight expansions, and
measuring noise correlations or mass currents. In the following
sections we show that they reveal additional information which helps
to identify the respective states. We should note that some of our
results rely on Monte-Carlo simulations of small systems or rough
estimates of particle energies and thus they should be considered as
qualitative estimates only.


\subsection{\label{sec:tofexp}Time-of-flight expansion}

The time-of-flight expansion is a standard measurement tool for
ultracold atomic systems. All potentials (including the artificial
magnetic field) are turned off instantaneously and the atomic cloud
is allowed to expand freely for a certain time before measuring its
density distribution. Since for an optical lattice setup
interactions during the expansion can be neglected and the time can
be chosen long enough such that the cloud expands to several times
its initial size the result of this process is \cite{Bloch-2007}
\begin{equation}
  \varrho_m(\mathbf{X})=|\tilde
   w(\mathbf{X}m_0/\hbar t)|^2\sum_{\mathbf{x}_1,\mathbf{x}_2}
   \varrho_1(\mathbf{x}_1;\mathbf{x}_2)
   \eh^{\ii\mathbf{X}\cdot(\mathbf{x}_1-\mathbf{x}_2)m_0/ \hbar t} \,.
\end{equation}
Here $\tilde w$ is the Fourier transform of the Wannier function
describing an atom in a single lattice site, $\varrho_1$ is the
one-particle density matrix, the sum is taken over all lattice site
vectors $\mathbf{x}_i$, $\varrho_m(\mathbf{X})$ is the measured
density at position $\mathbf{X}$ and time $t$ after release, and
$m_0$ is the free mass. Note that the ``momentum'' measured by this
process is the free space momentum $-i \hbar \nabla$. It is not to
be confused with the momentum of the system Hamiltonian $(-i \hbar
\partial_x - 2m\Omega y,  -i \hbar \partial_y)$. Furthermore, the
operators $\hat a_j$ need to be transformed into the laboratory
frame, and $\varrho_1$ is expressed in terms of these laboratory
frame operators. This is because the sudden switch-off of the
artificial magnetic field does not obey Maxwell's equations, which
breaks gauge invariance.

As shown in \cite{MacDonald-PRB-1988}, for a low $\alpha$ one can
calculate the single particle density matrix $\varrho_1$ of the
constant density lowest Landau level states, which include the
Laughlin, Read-Rezayi, 221, and NASS states. For large $\alpha$,
similar expressions with an additional factor $\sum_k{\mathbf
v}^{(k)*}{\mathbf v}^{(k)}$ from the small scale structure apply.
This gives for the density after the release
\begin{eqnarray}
\varrho_m(\mathbf{X})&=&|\tilde w(\mathbf{X}m_0/ \hbar
t)|^2\sum_{i,j}\sum_k{\mathbf v}^{\prime(k)*}(i){\mathbf
v}^{\prime(k)}(j)\frac{\tilde\nu}{2\pi n r_0^2}\nonumber\\&& \times
\exp\Bigg\{-\frac{(\mathbf{x}_i-\mathbf{x}_j)^2}{4r_0^2}\nonumber\\&&
  \quad + \ii \left(\frac{Xm_0}{\hbar t}+\frac{y_i+y_j}{4r_0^2}\right)(x_i-x_j)\nonumber\\
  &&\quad + \ii \left(\frac{Ym_0}{\hbar t}-\frac{x_i+x_j}{4r_0^2}\right)(y_i-y_j)\Bigg\} \,,
\end{eqnarray}
where the vectors $\mathbf{x}_j = (x_j,y_j)$ label the lattice
sites, and the prime indicates that we have taken the ${\mathbf
v}^{\prime (k)}$ in symmetric gauge, which doubles their period to
$2n$. Thus the arguments of the ${\mathbf v}^{\prime (k)}$ have to
be taken $\mathrm{modulus}\, 2n$.

Summing first over $\mathbf{x}_1-\mathbf{x}_2$ and approximating
this by an integral, we see that each $\mathbf{x}_1+\mathbf{x}_2$
contributes an approximate Gaussian centered on
$(-y_1-y_2,x_1+x_2)\hbar t/4m_0r_0^2$, and then summing over
$\mathbf{x}_1+\mathbf{x}_2$ we get the in-trap density profile
smeared out on a scale $r_0$, rotated through a right angle and
scaled by $\hbar t/4m_0r_0^2$.  Physically this happens because
lowest Landau level wavefunctions are very similar to rigid body
rotation, and once released from the field the atoms fly apart as if
they were rotating. Because the sum is in fact a discrete one, the
pattern repeats every reciprocal lattice cell (size $2\pi \hbar
t/dm_0$) with a decaying envelope $\tilde w$, and high $\alpha$
states can have multiple peaks per reciprocal lattice cell from the
small scale structure.

\begin{figure}
\includegraphics[width=8cm]{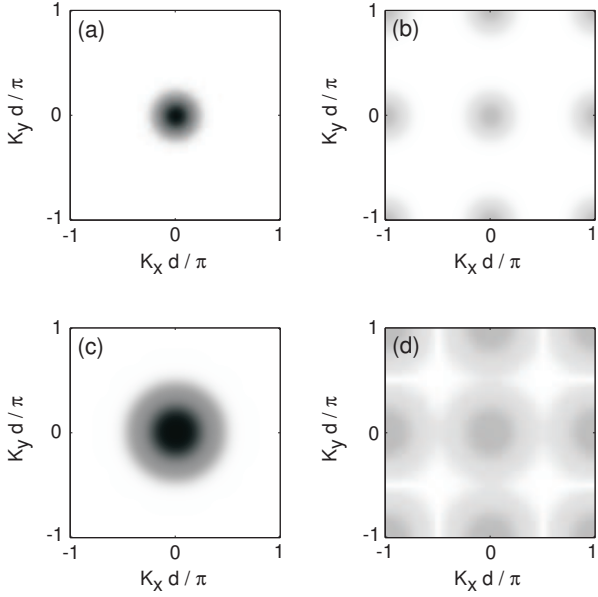}
\caption{\label{fig:tofexp-symm}Numerically calculated
time-of-flight expansions in symmetric gauge of (a,c) a Laughlin and
Read-Rezayi state ($\nu=1/2,1$), and (b,d) a 221 and NASS state
($\tilde\nu=2/3,4/3$). The initial states are circular with diameter
100 lattice sites for the outer ring and 50 for the inner one. The
artificial magnetic field is chosen such that (a) $\alpha=0.005$,
(b) $\alpha=0.505$, (c) $\alpha=0.01$, and (d) $\alpha=0.51$. The
resulting larger fields $\Omega$ or $\tilde \Omega$ in (c,d),
respectively, lead to a wider expansion of the cloud. The shown
expansions do \emph{not} include the $\tilde w$ slow decay, as the
speed of this is implementation dependent. A darker color
corresponds to a higher density of particles, and figures (a) and
(b) are to the same brightness scale, as are (c) and (d). Shown is
the first Brillouin zone, and $\mathbf{K} = m_0 \mathbf{X}/\hbar
t$.}
\end{figure}

\figref{fig:tofexp-symm} shows some numerically calculated examples,
the pattern repeats every reciprocal lattice cell with a slow decay
given by $\tilde w$ \cite{Greiner-Nature-2002}. As expected from the
above, fractional quantum Hall states have stepped peaks, further
distinguishing them from the featureless $\tilde w$ of a Mott
insulator \cite{Foelling-Nature-2005}, while vortex lattice states
have Thomas-Fermi (inverted parabola) peaks. Low $\alpha$ states
have one peak per reciprocal lattice cell while $\alpha\approx 1/2$
states have 4. The extra peaks (previously described for the vortex
lattice in \cite{Polini-LP-2004,Polini-PRL-2005}) confirm the
existence of a small scale structure. However, they do not show
whether one or more ``layers'' are occupied for multi-``layer''
systems.

As a time-of-flight expansion only measures single particle
properties, it cannot detect correlations, so different fractional
quantum Hall states are distinguished only by their density.  In the
next section, we will consider an enhancement of this method that
can measure the correlations of the states as well.


\subsection{\label{sec:correlation}Correlation measurement}

\begin{figure}
\centering
\includegraphics[width=6cm]{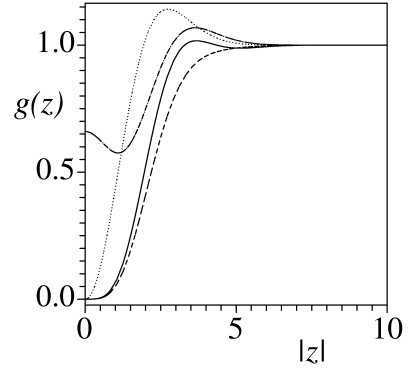}
\caption{\label{fig:correlator}Two-point functions for continuum
fractional quantum Hall states: $\nu=1/2$ Laughlin state (solid),
$\tilde\nu=2/3$ 221 state with $g_{11}=g_{22}$ (dashed) and
$g_{12}=g_{21}$ (dotted), and $\nu = 1$ Read-Rezayi state
(dash-dotted).}
\end{figure}

The two-particle correlations of a quantum state manifest themselves
in shot noise correlations of the density distribution after a
time-of-flight measurement \cite{Altman-PRA-2004,Read-PRA-2003}, as
was recently demonstrated experimentally for the Mott insulator
\cite{Foelling-Nature-2005}. At time $t$ after release from an
optical lattice these correlations are given by
\cite{Altman-PRA-2004}
\begin{equation}\label{Eq:Correlations}
\begin{split}
G({\mathbf{r,r^\prime}})\propto& \, t^{-6}\sum\limits_{ii^\prime
jj^\prime}\eh^{\ii{\mathbf
R}_{ii^\prime}\cdot{\mathbf{Q(r)}}+\ii{\mathbf
R}_{jj^\prime}\cdot{\mathbf{Q(r^\prime)}}}\av{\hat a^\dagger_i
\hat a^\dagger_j \hat a_{j^\prime} \hat a_{i^\prime}} \\
&-\av{ \hat n({\mathbf r})}\av{\hat n({\mathbf{r^\prime}})}\,,
\end{split}
\end{equation}
where $i$, $j$, $i^\prime$, $j^\prime$ run over all lattice sites,
${\mathbf R}_{ii^\prime}$ is the displacement vector from site $i$
to site $i^\prime$, and ${\mathbf{Q(r)}}=m_0{\mathbf r}/\hbar t$
with $m_0$ the free mass.  For states described by a continuum
wavefunction the expectation values over the quantum operators are
given by $\av{\hat a^\dagger_i \hat a^\dagger_{j} \hat a_{i'} \hat
a_{j^\prime}}=d^2\varrho_2(i,j;i',j')$. Here, $\varrho_2(i,j;i',j')$
is the continuum two-particle density matrix, which for lowest
Landau level states with constant density is of the form
\cite{MacDonald-PRB-1988}
\begin{equation}
\begin{split}\label{Eq:rho2_est}
  \varrho_2(i,j;i',j')=&\,\eh^{-\frac{1}{4}[|z_i|^2+|z_{i'}|^2+
  |z_j|^2+|z_{j'}|^2
-2(z_i^*z_{i'}+z_j^*z_{j'})]} \\
&\times\left(\frac{\tilde \nu}{2\pi n r_0^2}\right)^2
g((z_i-z_j)^*(z_{i'}-z_{j'})) \,.
\end{split}
\end{equation}
Note that $\tilde \nu = \nu$ and $n=1$ for $\alpha_c = 0$. The
function $g(z)$ for arbitrary complex $z$ is derived by analytic
continuation of the ordinary two-point correlation function
$g(|z_i-z_j|^2)$. If the behavior of a two-component state close to
$\alpha = 1/2$ is to be investigated, we need to replace
\begin{equation}
\begin{split}
  \av{\hat a^\dagger_i \hat a^\dagger_{j} \hat a_{i'} \hat
a_{j^\prime}} =& d^2 \sum_{k_1, k_2} \varrho_2^{(k_1,k_2)}
(z_i,z_j;z_{i'},z_{j'})
   \\
    &\qquad \times\mathbf{v}^{(k_1) \ast}(i ) \mathbf{v}^{(k_1)}(i')
  \mathbf{v}^{(k_2) \ast}(j ) \mathbf{v}^{(k_2)}(j')\,,
\end{split}
\end{equation}
where $\varrho_2^{(1,1)} = \varrho_2^{(2,2)}$ is the continuum
density matrix for two particles of the same type and
$\varrho_2^{(1,2)} = \varrho_2^{(2,1)}$ for two particles of
opposite type. The density matrices $\varrho_2^{(k_1,k_2)}$ are
described by expressions analogous to Eq.~(\ref{Eq:rho2_est}), where
the function $g(|z|^2)$ is replaced by $g_{k_1,k_2}(|z|^2)$. To
obtain $g_{k_1,k_2}$ we fit the series
\begin{gather}
  \begin{split}
  &g_{11}(|z|^2) = g_{22}(|z|^2) \\
  &\qquad= 1+\eh^{-|z|^2/2}+\sum_{m \geq 0\text{ even}}
     \frac{2c_m}{m!} \left(\frac{|z|^2}{4}\right)^m \eh^{-|z|^2 /4} \,,
  \end{split} \\
  g_{12}(|z|^2) = g_{21}(|z|^2) = 1+\sum_{m \geq 0} \frac{2 \tilde
     c_m}{m!} \left(\frac{|z|^2}{4}\right)^m \eh^{-|z|^2 /4} \,,
\end{gather}
to Monte Carlo data for the respective state. This works for any
lowest Landau level state as shown in
\cite{Girvin-PRB-1984,Girvin-PRB-1986}, where we take a sign change
into account because our particles are bosons. Our calculations show
that the Laughlin and 221 states have $g(0) = 0$ (see
Fig.~\ref{fig:correlator}), meaning that particles cannot come
together and the interaction energy is hence zero, while the NASS
and Read-Rezayi states have $g(0) > 0$ and hence non-zero
interaction energy (the NASS state is not shown since it is
qualitatively similar to the Read-Rezayi state).

Eq.~(\ref{Eq:Correlations}) involves a sum over four complex
variables, which makes its direct numerical evaluation
computationally intensive. To simplify this problem, we integrate
over $\mathbf{r} + \mathbf{r}'$, which gives a delta function
setting $z_i+z_j - z_{i^\prime} -z_{j^\prime} = 0$, and removes all
explicit dependence on $z_i+z_j + z_{i^\prime} + z_{j^\prime}$ . The
resulting correlation, which (ignoring the $t^{-6}$ prefactor) is a
function of momentum difference $\Delta \mathbf{k} =
\mathbf{Q}(\mathbf{r}) -\mathbf{Q}(\mathbf{r'})$, can hence be
evaluated for an infinitely extended state by summing over just two
complex variables, $z_i-z_j$ and $z_{i^\prime}-z_{j^\prime} $.
However, this does not work for a finite size state because the
ranges of $z_i - z_j$ and $z_{i^\prime}-z_{j^\prime} $ then do
depend on $z_i+z_j + z_{i^\prime} + z_{j^\prime}$, and assuming an
infinite size state while necessarily summing over a finite range
introduces the possibility of numerical artifacts.

\begin{figure}
\includegraphics[width=8cm]{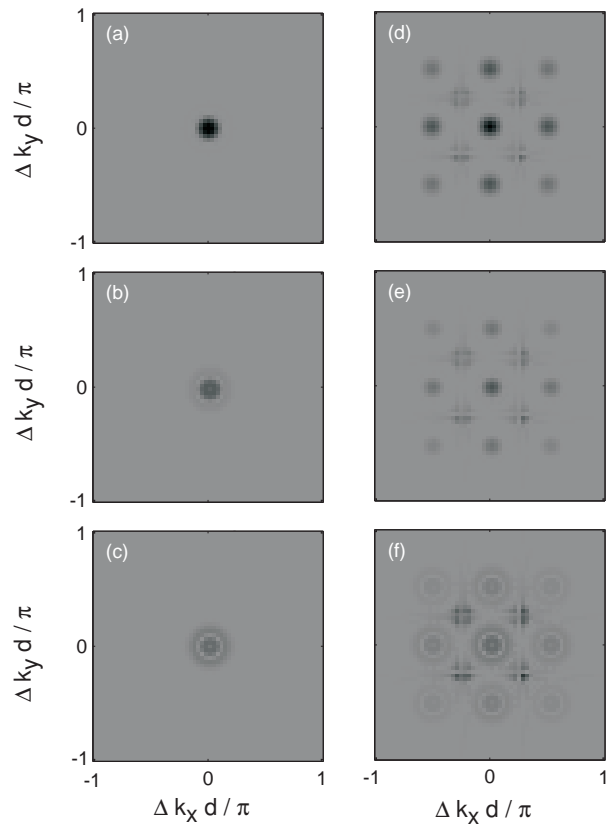}
\caption{\label{fig:corr-symm}Numerically calculated shot noise
correlations for an infinite system in symmetric gauge. The states
are (a) Laughlin, (b) $\nu=1$ Read-Rezayi, (c) $\nu=3/2$
Read-Rezayi, (d) 221, (e) $\tilde\nu=4/3$ NASS, and (f)
$\tilde\nu=3/2$ Read-Rezayi. The artificial magnetic field was
chosen to give $\alpha=0.01$ in (a)--(c), and $\alpha=0.51$ in
(d)--(f). Gray corresponds to no correlation, whereas black
corresponds to 100\% anticorrelation. We cannot rule out that the
four dips at $\Delta k=(\pm\pi /4d,\pm\pi /4d)$ are artifacts
arising from our approximations. }
\end{figure}

Some numerically calculated examples of the noise correlations are
shown in \figref{fig:corr-symm}, where again one reciprocal lattice
cell is shown and the pattern repeats. Fully Bose condensed states,
including ideal vortex lattices, have zero correlation.  The
Laughlin and 221 states are found to show near 100\% anticorrelation
at small $\Delta k$. Here we should stress that as this measurement
works in Fourier space, this anticorrelation is not the case for all
fully real-space anticorrelated states: the Mott insulator, for
example, shows positive correlation \cite{Foelling-Nature-2005}.
Higher density fractional quantum Hall states also show
anticorrelation but of reduced strength.  Our data suggests that
Read-Rezayi states have a ringed pattern with their strongest
anticorrelation ($\approx 50\%$ for $\nu=1$ and $\approx 35\%$ for
$\nu=3/2$) at a nonzero $\Delta k$, while the $\tilde\nu=4/3$ NASS
state has $\approx 40\%$ anticorrelation at zero $\Delta k$ and no
ring. The presence or absence of this ring might be used to
distinguish between the Read-Rezayi and NASS states. Apart from this
we observe that the small-scale structure for $\alpha \approx 1/2$
is also visible in the noise-correlations, see
Figs.~\ref{fig:corr-symm}(d)-\ref{fig:corr-symm}(f): Additional
anticorrelation dips at $\Delta k = (\pm \pi/2d,\pm \pi/2d ),\,(\pm
\pi/2d,0),\,(0,\pm \pi/2d)$ occur, which have a similar structure to
the central ones at $\Delta k = (0,0)$.

\subsection{\label{sec:dynamics}Hall current and disorder}

So far, we discussed only the static properties of the fractional
quantum Hall states and how to detect them. However, in solid state
physics one of the most important observables is a dynamical
property, namely the Hall current. In this section, we will
therefore study the atomic case subject to a linear tilt and to
disorder.

When a linear potential $V(x,y)=may$ is applied to a continuum
fractional quantum Hall system, all states acquire a velocity
$a/(2\Omega)$ at right angles to the potential gradient causing the
Hall current. In a perfect continuum system this is an exact result,
unaffected by interaction, because such an acceleration term is
exactly canceled out by Galilean transforming
Eq.~(\ref{Eq:ContAppIA}) to a reference frame moving with this
velocity. In a lattice system the result is valid at sufficiently
low $\alpha$ for the continuum approximation to apply; at higher
$\alpha$ the lattice, which defines a rest frame, becomes important
and a different velocity can occur \cite{Palmer-PRL-2006}.  In
particular, near simple rational $\alpha$ where
Eq.~(\ref{eq:chi-manypar}) is valid, the velocity is
$a/(2\tilde\Omega)$, which can be very different: for
$\alpha<\alpha_c$ it even has the opposite sign, meaning that the
current flows in the opposite direction. A negative Hall current was
also observed for a single particle in a lattice using Kubo response
theory \cite{Bhat-PRA-2007}.

For nonlinear potentials of large length scale compared to the
magnetic length $r_0$, the single particle eigenstates lie along the
equipotential lines of the potential and the Hall current flows
along those lines at the velocity given by the local potential
gradient. This motion is not visible in equilibrium as the steps in
the density profile lie along equipotentials as well. However, it
can be made visible by putting the system out of equilibrium, for
example, by suddenly changing the trapping potential. Relaxation to
equilibrium will be slow because nonforward scattering is
energetically forbidden in the fractional quantum Hall system,
making the Hall current a supercurrent
\cite{Prange-1990,Yennie-RMP-1987}.

It has been shown that when sufficiently mild disorder is added to a
fractional quantum Hall system, some of the particles become
localized and cannot carry current, but those which remain free move
faster and the average velocity is still $a/(2\tilde\Omega)$
\cite{Prange-1990,Yennie-RMP-1987}. For example, for smooth disorder
$V(x,y) = may + V_r(x,y)$, where $V_r$ is a random potential with
length scale much larger than $r_0$ and zero average, the expression
for the $x$ direction of the velocity, $v_x = (\partial V/\partial
y)/(2m\tilde\Omega)$, remains valid. Hence particles on closed
equipotentials are confined to those lines, but the random part of
$v_x$ averages to zero so the average velocity in the $x$ direction
remains $a/(2\tilde\Omega)$. The current is hence determined by the
width over which each extended level is occupied. A simple model of
this is to describe the disorder by a density of states
$\varrho_d(E)$, given by the number of lowest Landau level states
per unit energy interval and per unit area. This gives $\int
\varrho_d(E) \,\diff E = m\tilde\Omega/(\pi \hbar)$. In the case of
smooth disorder, $\varrho_d(E)$ is proportional to the probability
distribution of the noise potential. For a linear geometry, we then
have
\begin{gather}
  N = L \int \! \diff y \int\limits^{\mu-V_1(y)} \! \diff E
      \, \sum_{j\geq 0}(\nu_j-\nu_{j-1})\varrho_d(E-\mu_j) \,,\\
  I = \frac{m}{2\pi\hbar} \int \! \diff y \, \frac{d V_2}{dy}
  \int\limits^{\mu-V_1(y)} \! \diff E \, \sum_{j\geq0}
      (\nu_j-\nu_{j-1})\delta(E-\mu_j)\,,
\end{gather}
where $L$ is the length of the system along the $x$ axis, $I$ is the
net current in the $x$ direction, $N$ the number of particles, and
$\mu$ the chemical potential. The fractional quantum Hall states are
taken at filling factors $\nu_j$ and chemical potential $\mu_j$, and
$V$ is suddenly changed from $V_1(y)$ to $V_2(y)$ to achieve the
nonequilibrium situation. The exact shape of the disorder and thus
of $V_1(y)$ and $V_2(y)$ are not important for our qualitative
investigations. These equations are valid for weak trapping
potentials, where $dV/dy$ is much smaller than the disorder term.
Stronger potentials can break weakly localized states free,
replacing the $\delta(E-\mu_j)$ density of extended states in $I$ by
a finite width distribution.

\begin{figure}[t!]
\centering
\includegraphics[width=7cm]{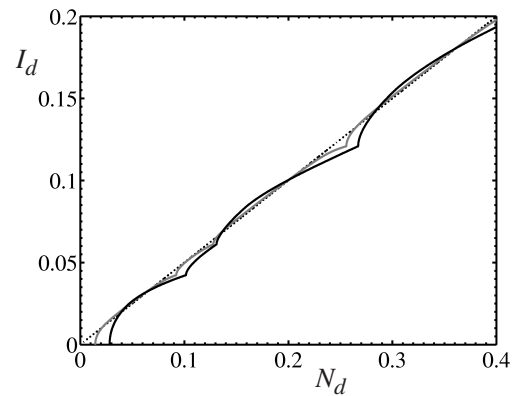}
\caption{\label{fig:current}
Dimensionless Hall current $I_d = I/(2 a m^2\sqrt{2 u
\tilde\Omega/\hbar^3}/\omega )$ against dimensionless number of
atoms per unit length $N_d = N/( 2 L m \sqrt{2 u
\tilde\Omega^3/\hbar^3}/\omega)$, where $L$ is the length of the
system, which is contained in a 1D harmonic trap $V (x, y) = m
\omega^2 y^2/2 + yma$ at $ \alpha \approx 1/2$. The linear term
$yma$ is turned on after the atoms have come to equilibrium in the
trap, and the curves are for no disorder (dotted straight line),
maximal disorder (constant density of states, black curve), and
Lorentzian disorder of width $(1/5) u m \tilde \Omega/2 \pi \hbar $
(gray curve). The corresponding plot for small $\alpha$ is
qualitatively similar \cite{Palmer-PRL-2006}. }
\end{figure}

In a square-well potential (as in a solid state system) the
localized states create a finite range of filling factors over which
a given number of fractional quantum Hall extended state levels are
completely full, giving rise to the almost perfectly flat fractional
quantum Hall plateaus. In a harmonic trap there will not be
fractional quantum Hall plateaus, only corners each time a new
extended level begins to fill, see Fig. \ref{fig:current}. Between
those corners, there exist different fractional quantum Hall states.
Unlike the square-well case, it is possible to obtain the complete
distribution $\varrho_d(E)$ by measuring $I$ against $N$ (or
$\tilde\Omega$).


\section{\label{sec:fqh-conclusions}Conclusions}

In our work we have studied the low and high field fractional
quantum Hall effect in an optical lattice. For small $\alpha$,
corresponding to low field strengths, the continuum approximation
can be employed to find the solutions of the Hamiltonian. For high
fields corresponding to values of $\alpha$ close to simple
rationals, $\alpha \approx \alpha_c = l/n$, we showed that the
states can be approximated by an $n$-component wave function, where
each component is described by a slowly varying, continuous function
$\chi^{(k)}$. These functions are solutions of a differential
equation similar to the $\alpha \ll 1$ case, but with an artificial
magnetic field term $\tilde \Omega = \hbar \pi (\alpha - \alpha_c)/m
d^2$ and an effective filling factor $\tilde\nu \equiv \hbar
\varrho\pi/m\tilde\Omega$ instead of the usual $\Omega$ and $\nu$.

We have shown that many interesting topological states may occur in
our setup, even if it is subject to a weak harmonic trapping
potential in one direction. These states include the Laughlin, the
Read-Rezayi, the 221, and the non-Abelian spin singlet (NASS) state.
Numerical investigations also showed the occurrence of a striped
vortex lattice phase close to $\alpha \approx 1/2$, as predicted for
a two-component system at \emph{low} $\alpha$. In our case the two
components did not correspond to two different atomic species, but
to the two components $\chi^{(k)}$ of the wave function.

We have furthermore demonstrated that the above states can be
distinguished by suitable measurements. Simple time-of-flight
expansions reveal additional structures for higher $\alpha$ states,
but do not necessarily allow for distinguishing between the
different quantum Hall states. More insight can be gained by
measuring two-point correlation functions. For the Laughlin state,
this function is zero for $|z_1-z_2|=0$, meaning that there are
never two atoms at the same place, whereas for the Read-Rezayi state
with $\nu = 1$ the correlation function exhibits a finite value at
zero distance. We, moreover, showed that the different states also
exhibit different noise correlations, which can be employed to
distinguish between them. Again, higher $\alpha$ states show
additional structures due to their additional small scale
symmetries. Further, our results suggest that the Read-Rezayi states
exhibit a ring structure, which allows one to distinguish them from
the Laughlin or 221 states.

In analogy to the semiconductor Hall effects, it is possible to
displace the optical lattice and measure mass transport effects,
which behave similarly to the condensed matter analogues. Especially
Hall currents proportional to $\alpha-\alpha_c$ rather than $\alpha$
have been predicted, which may lead to a negative sign in the
direction of the mass flow.


\acknowledgments{The authors thank Sarah Al-Assam for helpful
comments. This work was supported by the United Kingdom EPSRC
through QIP IRC (Grant No.~GR/S82176/01) and EuroQUAM Project
No.~EP/E041612/1, the EU through the STREP project OLAQUI, the
Merton College Domus Scholarship (R.N.P.), and the Keble Association
(A.K.).}

\medskip


\appendix

\section{\label{app:high-field}Derivation of high $\alpha$ states}

In this appendix we give details on the derivation of the
Hamiltonian equation (\ref{eq:chi-1par}) for large $\alpha$ close to
a rational $\alpha_c = l/n$ with small integers $l$ and $n$. We
assume that apart from the optical lattice the atomic gas
experiences no trapping potential in the $x$ direction, but might be
trapped by a (slowly varying) harmonic potential $V(x,y)
=\frac{1}{2}m\omega^2y^2$ in the $y$-direction. Note that the mass
in this potential is the effective mass $m = \hbar / 2 J d^2$
introduced earlier. Motivated by previous findings
\cite{Palmer-PRL-2006}, we make for the wave function the ansatz
$\psi(ns+i,nr+j)=d \phi_j((nr+j)d)\eh^{\ii K(ns+i)d}$. Here, $i,j =
1,...,n$ and $r,s$ are integers, and we define $x = (ns + i)d$, $y =
(nr+j)d$. Since $\hat H$ is independent of $x$ if the potential $V$
is, the $x$ quasimomentum $K$ is conserved exactly for our choice of
$V$. Let $K = 2\pi k l /nd+ \tilde K$ , where $k$ is an integer and
$\tilde K \ll 1/n$ to yield an $n$ site periodicity plus slow
variation in the $x$ direction. The action of Hamiltonian equation
(\ref{Eq:Hamiltonian}) with $U=0$ gives
\begin{widetext}
\begin{equation} \label{Eq:App1}
\begin{split}
  \hat H \ket{\psi} =& - J d \sum_{r,s} \sum_{i,j} \bigg[
    \eh^{2 \pi \ii \alpha (nr+j) + \ii K (ns+i-1)d} \phi_j(nr+j)
    +\eh^{-2 \pi \ii \alpha (nr+j) + \ii K (ns+i+1)d}
    \phi_j(nr+j)\\
    &\qquad + \phi_{j-1}(nr+j-1) \eh^{\ii K (ns+i)d}
            + \phi_{j+1}(nr+j+1) \eh^{\ii K (ns+i)d}\bigg] \hat
            a^\dagger_{nr+j,ns+i} \ket{\mathrm{vac}} \\
            & + \sum_{r,s} \sum_{i,j} V(nr+j,ns+i) \hat
            a^\dagger_{nr+j,ns+i} \ket{\mathrm{vac}} \,.
\end{split}
\end{equation}
Using the fact that $\phi_j$ is only slowly varying for constant
index $j$ \cite{Palmer-PRL-2006}, we can approximate derivatives by
appropriate discrete differences and make use of the fact
\begin{equation}
  \frac{\phi_{j \pm 1}(y)}{d^2} \pm
  \frac{1}{d} \frac{\partial \phi_{j\pm 1} (y)}{\partial y}
  + \frac{1}{2} \frac{\partial^2 \phi_{j \pm 1} (y)}{ \partial y^2}
  \approx
  \frac{1}{d^2} \phi_{j\pm 1} [(nr + j \pm 1)d] \,.
\end{equation}
Collecting all terms with the same creation operators in
Eq.~(\ref{Eq:App1}) we can derive a Hamiltonian for $\phi_j$, acting
as
\begin{equation}
\begin{split}
  \hat H_1 \phi_j =& -\frac{\hbar^2}{md^2}\cos\left(\frac{2\pi j l}{n}+  \frac{2m\tilde\Omega
y d}{\hbar} -K d \right) \phi_j + \frac{1}{2} m \omega^2 y^2 \phi_j
-\frac{\hbar^2}{2m} \left(\frac{\phi_{j+1}}{d^2}+\frac{1}{d}
\frac{\partial\phi_{j+1}}{\partial y}+\frac{1}{2}
\frac{\partial^2\phi_{j+1}}{\partial y^2}
\right.\\
&\quad\left.+\frac{\phi_{j-1}}{d^2}-\frac{1}{d}\frac{\partial\phi_{j-1}}{\partial
y}+\frac{1}{2}\frac{\partial^2\phi_{j-1}}{\partial y^2}\right)+O(d)
\,.
\end{split}
\end{equation}
If the condition $2 m \tilde \Omega y d /\hbar \ll 1$ holds over the
range where the wave function is appreciable, namely $y \sim
l_\chi$, we can apply the Taylor expansion to the cosine, which
after collecting terms in orders of $d$ yields
\begin{equation}
\begin{split}
  \hat H_1 \phi_j =& -\frac{\hbar^2}{2md^2}\left[\phi_{j+1}+2\cos\left(\frac{2\pi
   (j-k)l}{n}\right)\phi_j+\phi_{j-1}\right]-
   \frac{\hbar^2}{2md}\left[(\tilde K-\frac{2m\tilde\Omega y}{\hbar})
   2\sin\left(\frac{2\pi (j-k)l}{n}\right)\phi_j
   +\frac{\partial\phi_{j+1}}{\partial y}
   -\frac{\partial\phi_{j-1}}{\partial y}\right]   \\
   &-\frac{\hbar^2}{2m} \left[ -\left(\tilde K-\frac{2m\tilde\Omega y}{\hbar}\right)^2
   \cos\left(\frac{2\pi (j-k)l}{n}\right)\phi_j -
    \frac{m^2\omega^2y^2}{\hbar^2}  \phi_j
    +\frac{1}{2}\frac{\partial^2\phi_{j+1}}{\partial y^2}
    +\frac{1}{2}\frac{\partial^2\phi_{j-1}}{\partial
    y^2}\right]+O(d) \,.
\end{split}
\end{equation}
Note that for $n=1$ or 2 the odd powers in $d$ cancel by symmetry
and we get an expansion in $d^2$, but that for larger $n$ the
expansion is in $d$. Define $\Phi=(\phi_1\ldots\phi_n)^T$ and expand
in powers of $d$: $\Phi= {\Phi}^{(0)} + d{\Phi}^{(1)}+ d^2{
\Phi}^{(2)}+O(d^3)$. Similarly expand the energy $E = E_0/d^2 +
E_1/d + E_2 +O(d)$. We define the matrix
\begin{equation}
    {\mathbf A}_0 = \left(\begin{array}{cccccc}
       2\cos 2\pi (1-k) l/n &1&0&\cdots&0&1\\
       1  &2\cos 2\pi (2-k) l/n&1&&&0\\
       0&1&\ddots&\ddots&&\vdots\\
       \vdots&&\ddots&\ddots&\ddots&0\\
       0&&&\ddots&\ddots&1\\
       1&0&\cdots&0&1&2\cos 2\pi(n-k) l/n
       \end{array}\right) \,.
\end{equation}
\end{widetext}
The $O(1/d^2)$ terms then become ${\mathbf A}_0{ \Phi}^{(0)} = -2 m
E_0 {\Phi}^{(0)}/\hbar^2$, so ${\Phi}^{(0)}$ is an eigenvector of
${\mathbf A}_0$, with the ground state having the largest
eigenvalue. Assuming nondegenerate eigenvalues, $\Phi^{(0)}(y)$ is
hence proportional to the same (normalized) eigenvector ${\mathbf
v}^{(k)}$, and can depend on $y$ only in overall magnitude, i.e.,
${\Phi}^{(0)}(y)=\chi(y){\mathbf v}^{(k)}$.

Since changing $k$ is equivalent to changing the origin of $j$, the
eigenvalues are the same for all $k$, with ground state eigenvectors
$v^{(k)}_j = v_{j-k}$, where $\mathbf{v}$ is the normalized ground
state eigenvector for $k = 0$ and the subscript $j - k$ wraps around
modulus $n$. This gives $n$ degenerate ground states $k = 0,...,
n-1$, which are orthogonal because of their different $K$ values.
For simplicity we take $k = 0$ in the remainder of this derivation.

For the $O(1/d)$ terms, define
\begin{equation}
  {\mathbf A}_1=\left(\begin{array}{ccccc}
  0&1&0&\cdots&-1\\
  -1&0&1&&\vdots\\
  0&-1&0&\ddots&0\\
   \vdots&&\ddots&\ddots&1\\
   1&\cdots&0&-1&0
   \end{array}\right) \,,
\end{equation}
\begin{equation}
  {\mathbf{A}}_2= 2\mathrm{diag}[\sin (2\pi l/n),\sin( 4\pi l/n),...,\sin( 2n \pi
  l/n)] \,,
\end{equation}
where $\mathrm{diag}$ denotes a diagonal matrix with the argument as
the entries on the diagonal. Define furthermore ${\mathbf
w}_1={\mathbf A}_1{\mathbf v}$, ${\mathbf w}_2={\mathbf A}_2{\mathbf
v}$. This yields
\begin{equation}
\begin{split}\label{eq:phi1}
   ({\mathbf A}_0+2mE_0/\hbar^2){\mathbf \Phi}^{(1)}(y)=&\,\frac{2mE_1}{\hbar^2}
   \chi(y){\mathbf v}
   +\frac{d\chi}{dy}{\mathbf w}_1\\
   &\quad +\left(K-\frac{2m \tilde\Omega y}{\hbar} \right)
   \chi(y){\mathbf w}_2   \,.
\end{split}
\end{equation}
The left hand side of Eq.~(\ref{eq:phi1}) is orthogonal to ${\mathbf
v}$ because ${\mathbf A}_0+2mE_0/\hbar^2$ annihilates ${\mathbf v}$
and is Hermitian, while ${\mathbf w}_1$ is orthogonal to ${\mathbf
v}$ because ${\mathbf A}_1$ is antisymmetric and hence so is
${\mathbf v}^T{\mathbf A}_1{\mathbf v}$, but the latter is a number
so it can only be antisymmetric if it is zero.  Hence a solution can
only exist if ${\mathbf w}_2$ is also orthogonal to ${\mathbf v}$
and $E_1=0$; this is the case for all $\alpha_c =
1/8,1/7,1/6,1/5,1/4,2/7,1/3,3/8,2/5,3/7,1/2$ within numerical
accuracy \cite{Palmer-PRL-2006}, but we have not been able to prove
that it is always the case.

For the $O(1)$ terms define
\begin{equation}
  {\mathbf A}_3=\left(\begin{array}{ccccc}
    0&1&0&\cdots&1\\
    1&0&1&&\vdots\\
    0&1&0&\ddots&0\\
    \vdots&&\ddots&\ddots&1\\
    1&\cdots&0&1&0
    \end{array}\right) \,,
\end{equation}
\begin{equation}
  {\mathbf{A}}_4= \mathrm{diag}[\cos (2\pi l/n),\cos( 4\pi l/n),...,\cos( 2n \pi
  l/n)] \,,
\end{equation}
giving
\begin{equation}\label{Eq:App_order_3}
\begin{split}
  &\left(\frac{2mE_2}{\hbar^2}-\frac{m^2\omega^2y^2}{\hbar^2}\right)\chi(y){\mathbf v}\\
  &-\left(\frac{2m\tilde\Omega y}{\hbar}-K\right)^2\chi(y){\mathbf A}_4{\mathbf v}
  + \frac{1}{2}\frac{\partial^2\chi}{\partial y^2}{\mathbf A}_3{\mathbf v}  \\
  &+ \left[\left(K-\frac{2m\tilde\Omega y}{\hbar}\right){\mathbf A}_2+
  \frac{\partial}{\partial y}{\mathbf A}_1\right]
  \left({\mathbf A}_0+\frac{2mE_0}{\hbar^2}\right)^{-1}  \\
  &\quad\times\left[\frac{\partial \chi}{\partial y}{\mathbf w}_1
  +\left(K-\frac{2m\tilde\Omega y}{\hbar}\right)\chi(y){\mathbf w}_2\right]  \\
  &+\left({\mathbf A}_0+\frac{2mE_0}{\hbar^2}\right){\mathbf \Phi}^{(2)}
  ={\mathbf 0} \,.
\end{split}
\end{equation}
We note that for nondegenerate eigenvalues of $\mathbf{A}_0$ the
expressions $(\mathbf{A}_0 + 2 m E_0/\hbar^2)^{-1} \mathbf{w}_{j}$,
$j = 1,2$, are well-defined since both $\mathbf{w}_1$ and
$\mathbf{w}_2$ are orthogonal to $\mathbf{v}$. Taking the scalar
product of Eq.~(\ref{Eq:App_order_3}) with ${\mathbf v}$ gives a
harmonic oscillator equation for $\chi(y)$,
\begin{equation}
  E_2 \chi =  -\frac{\hbar^2 C_1}{2m}\frac{\partial^2\chi}{\partial
  y^2}
  +\frac{\hbar^2 C_2}{2m}\left(\frac{2m\tilde\Omega
  y}{\hbar}-K\right)^2\chi
  +V(x,y)\chi \,,
\end{equation}
where the dimensionless constants $C_1={\mathbf v}^T{\mathbf
A}_3{\mathbf v}/2-{\mathbf w}_1^T({\mathbf
A}_0+2mE_0/\hbar^2)^{-1}{\mathbf w}_1$ and $C_2={\mathbf
v}^T{\mathbf A}_4{\mathbf v}-{\mathbf w}_2^T({\mathbf
A}_0+2mE_0/\hbar^2)^{-1}{\mathbf w}_2$ depend only on $l$ and $n$,
and by changing the roles of $x$ and $y$ by using a gauge
transformation one can show that $C_1 = C_2 \equiv C$. This
oscillator has mass $m/C_1$, frequency $\omega_{\mathrm{ eff}}=(4C_1
C_2 \tilde\Omega^2+ C_1 \omega^2)^{1/2}$, and center $y_c= 2\hbar
C_2 K \tilde\Omega/(4C_2m\tilde\Omega^2+m\omega^2)$.




\bibliography{../../Biblio}

\end{document}